\documentclass{aa}  

\usepackage{graphicx}
\usepackage{txfonts}

\usepackage[normalem]{ulem}

\begin{document}

\def\beginthetable{ \begin{table*} }
\def\endthetable{ \end{table*} }

\def\absPath{"./"}

   \title{Herschel/PACS\thanks{{\it Herschel} is an ESA space observatory with science instruments provided by European-led Principal Investigator consortia and with important participation from NASA.} observations of the 69 $\mu m$ band of crystalline olivine around evolved stars\thanks{The Appendix A  is only available in the on line version. The spectra are available in electronic form
at the CDS via anonymous ftp to cdsarc.u-strasbg.fr (130.79.128.5)
or via http://cdsweb.u-strasbg.fr/cgi-bin/qcat?J/A+A/}}


\author{J.A.D.L.~Blommaert \inst{1,2,11}
        \and B.L.~de~Vries \inst{1,12,13}
        \and L.B.F.M.~Waters \inst{3,4} 
        \and C.~Waelkens \inst{1}
        \and M.~Min \inst{4}
        \and H.~Van~Winckel \inst{1}
        \and F.~Molster \inst{5}
        \and L.~Decin \inst{1}
        \and M.A.T.~Groenewegen \inst{6}
        \and M.~Barlow \inst{7}
       \and P.~Garc\'ia-Lario \inst{8}
        \and F.~Kerschbaum \inst{9}
        \and Th.~Posch \inst{9}
        \and P.~Royer \inst{1}
        \and T.~Ueta \inst{10}
        \and B.~Vandenbussche \inst{1}
         \and G.~Van de Steene \inst{6}
        \and P.~van Hoof \inst{6}
}

\authorrunning{Joris Blommaert \& Ben de Vries et al.}

\institute{ Instituut voor Sterrenkunde, K.U. Leuven, Celestijnenlaan 200D, 3001 Leuven, Belgium
       \and Astronomy and Astrophysics Research Group, Department of Physics and Astrophysics, Vrije Universiteit Brussel, Pleinlaan 2, 1050 Brussels, Belgium
        \and SRON Netherlands Institute for Space Research, Sorbonnelaan 2, 3584 CA Utrecht, The Netherlands
        \and Sterrenkundig Instituut Anton Pannekoek, University of Amsterdam, Science Park 904, 1098 XH, Amsterdam, The Netherlands
        \and Leiden Observatory, Leiden University, PO Box 9513, 2300 RA, Leiden, The Netherlands
        \and Koninklijke Sterrenwacht van Belgi\"e, Ringlaan 3, 1180 Brussel, Belgium
        \and Department of Physics and Astronomy, University College London, Gower Street, London WC1E 6BT, UK
        \and Herschel Science Centre, European Space Astronomy Centre, Research and Scientific Support Department of ESA, Villafranca del Castillo, E-28080 Madrid, Spain 
        \and University of Vienna, Department of Astrophysics, T\"urkenschanzstrasze 17, 1180 Wien, Asutria
        \and Department of Physics and Astronomy, MS 6900, University of Denver, Denver, CO 80208, USA
       \and  Flemish Institute for Technological Research (VITO), Boeretang 200, 2400 Mol, Belgium (present address)
        \and AlbaNova University Centre, Stockholm University, Department of Astronomy, SE-106 91 Stockholm, Sweden
       \and Stockholm University Astrobiology Centre, SE-106 91 Stockholm, Sweden
}

   \date{Received 28 August 2013; accepted 10 February 2014}

 
  \abstract
   {We present 48 Herschel/PACS spectra of evolved stars in the wavelength range of 67-72 $\mu$m. This wavelength range covers the 69 $\mu$m band of crystalline olivine ($\text{Mg}_{2-2x}\text{Fe}_{(2x)}\text{SiO}_{4}$). The width and wavelength position of this band are sensitive to the temperature and composition of the crystalline olivine. Our sample covers a wide range of objects: from high mass-loss rate AGB stars (OH/IR stars, $\dot M \ge 10^{-5}$~M$_\odot$/yr), through 
post-AGB stars with and without circumbinary disks, to planetary nebulae and even a few massive evolved stars.}
  { The goal of this study is to exploit the spectral properties of the 69 $\mu$m band to determine the composition and temperature of the crystalline olivine. Since the objects cover a range of evolutionary phases, we study the physical and chemical properties in this range of physical environments.}
   {We fit the 69 $\mu$m band and use its width and position to probe the composition and temperature of the crystalline olivine.}
   {For 27 sources in the sample, we detected the 69 $\mu$m band of crystalline olivine ($\text{Mg}_{(2-2x)}\text{Fe}_{(2x)}\text{SiO}_{4}$). The 69 $\mu$m band shows that all the sources produce pure forsterite grains containing no iron in their lattice structure. The temperature of the crystalline olivine as indicated by the 69 $\mu$m band, shows that on average the temperature of the crystalline olivine is highest in the group of OH/IR stars and the post-AGB stars with confirmed Keplerian disks. The temperature is lower for the other post-AGB stars and lowest for the planetary nebulae. A couple of the detected 69 $\mu$m bands are broader than those of pure magnesium-rich crystalline olivine, which we show can be due to a temperature gradient in the circumstellar environment of these stars. The disk sources in our sample with crystalline olivine are very diverse. They show either no 69~$\mu$m band, a moderately strong band, or a very strong band, together with a temperature for the crystalline olivine in their disk that is either very warm ($\sim$600~K), moderately warm ($\sim$200~K), or cold ($\sim$120~K), respectively.}
   {}

   \keywords{Stars: AGB and post-AGB -- Stars:  outflows -- circumstellar matter -- dust 
               }

   \maketitle
%


\section{Introduction}
Crystalline olivine ($\text{Mg}_{(2-2x)}\text{Fe}_{(2x)}\text{SiO}_{4}$) has been detected in many different circumstellar environments, like disks around pre-main-sequence stars \citep{waelkens96, meeus01, sturm13}, comets \citep{wooden02}, post-main-sequence stars \citep{waters96, mol02_3}, and active galaxies \citep{spoon06, kemper07}. In outflows around evolved stars, the features of crystalline silicates were first detected in the mid-infrared wavelength range. For crystalline olivine, the most prominent bands in this range are at 11.3, 23.6, and 33.6 $\mu$m, which come from the Si-O-Si or O-Si-O bending modes and from the translational and rotational dominated modes, respectively. Interaction between the cations and anions make these bands sensitive to the ratio Mg/Fe in the mineral \citep{koike03}. With this sensitivity it was shown that the iron content of crystalline olivine in circumstellar environments of evolved stars is lower than 10\% \citep{tielens98, mol02_3}.

At longer wavelengths than 45 $\mu$m, the spectrum of crystalline olivine shows resonances around 49 and 69~$\mu$m \citep{bowey01, bowey02, koike03, koike06, suto06}. The wavelength position and width of this feature strongly depend on both the Mg/Fe ratio of the mineral, as well as on the grain temperature. The 69~$\mu$m band is isolated and sits on top of a smooth continuum, allowing for a very precise analysis of the feature. With the Herschel/PACS spectrometer \citep{pilbratt10, poglitsch10}, operating in the 50--200~$\mu$m range,  it is now possible to study the 69~$\mu$m band for a broad range of targets.  Such an analysis has already been done for young stellar objects, showing that the crystalline olivine in proto-planetary disks and debris disks is very magnesium-rich ($<$0-3\% iron, \cite{mulders11, devries12, sturm10, sturm13}). 

In this paper we describe our study of the forsterite dust observed in a sample of evolved stars of predominantly  low to intermediate initial mass ($\lessapprox 8 $M$_{\odot}$). In the final stages of a stellar life, the evolution is dominated by strong mass loss. During the asymptotic giant branch (AGB) phase large amplitude pulsations bring the gas to high enough altitudes for dust particles to condense. Radiation pressure on the dust and drag between the dust and gas create a strong and dense outflow. The mass-loss rates of AGB stars can be in the range of $10^{-8}$ \nolinebreak M$_{\odot}$ \nolinebreak yr$^{-1}$ up to $10^{-4}$ M$_{\odot}$ yr$^{-1}$ \citep{HO03}. 

Depending on the oxygen and carbon abundance in the gas, either an oxygen- or carbon-rich chemistry is initiated in the outflow, because the less abundant element will be locked into the CO molecule. In the case of an oxygen-rich outflow (C/O$<$1), oxygen-rich dust such as 
olivine and pyroxene ($\text{Mg}_{1-x}\text{Fe}_{x}\text{SiO}_{3}$) is formed. In the carbon-rich case, carbon dust particles are formed. While the exact values are under debate and are certainly metal dependent, oxygen-rich stars with initial masses between 1.5 $\text{M}_{\odot}$ and 3-4 $\text{M}_{\odot}$ can transition into a carbon-rich star  
through dredge-up of sufficient carbon from the core to the outer layers of the star. More massive oxygen-rich stars of $\ge$3-4 $\text{M}_{\odot}$ will remain oxygen-rich because the hot bottom burning in these objects favours the production of nitrogen instead of carbon \citep{HO03}.

When the AGB star is stripped of most of its outer layers, its strong pulsations cease, and the mass loss stops. In this case the star goes through a short post-AGB or proto-planetary nebula (proto-PN) phase, where the previously emitted material slowly drifts away from the star and cools, while the central stars heat up \citep{HO03}. Eventually the core becomes visible as a white dwarf, signaling the end of stellar evolution. The white dwarf briefly ionizes the emitted material, creating a PN. 

Studies based on thermodynamic equilibrium have shown that crystalline olivine is a mineral that is expected to form around 1400K to 1000 K if a cloud of gas with solar abundance cools \citep{sedlmayr89, tielens98}. Around these temperatures, crystalline olivine forms as its magnesium-rich end-member 
forsterite ($\text{Mg}_{2}\text{SiO}_{4}$). These magnesium-rich crystalline olivine grains are indeed observed to be moderately abundant in the outflow of AGB stars (2-12\% of the total dust mass \citep{devries10}). If enough SiO is present at a temperature of 1300K to1000 K, it can react with forsterite to form enstatite (MgSiO$_{3}$), the magnesium-rich end-member of the pyroxene group. Subsequent reactions with gas phase iron converts enstatite to iron containing crystalline olivine around 1100 K to 900 K. In many circumstellar environments of evolved stars, enstatite is indeed detected \citep{mol02_3}, and Fe is likely to be present (although it is not certain whether it is present in the gas phase). Based on thermodynamic equilibrium, iron containing crystalline olivine could thus be formed. However, if the densities become too low in the cooling outflows of evolved stars, thermodynamic equilibrium does not need to hold, and reactions are expected to "freeze out". 

Gas-phase condensation is not the only possible formation mechanism of crystalline olivine. An alternative is to form crystalline olivine by annealing amorphous olivine. When amorphous olivine is heated above the glass temperature of olivine, it can rearrange its lattice elements into a crystalline form.  \cite{sogawa99} show that amorphous silicate grains accreted on precondensed corundum grains can be crystalized for high enough mass loss rates ($> 3 \times 10^{-5}$ M$_{\odot}$ yr$^{-1}$ in the case of an L$_* = 2 \times 10^4$~L$_{\odot}$ star). In their study of the Spitzer IRS spectroscopy of oxygen-rich AGB stars, \cite{jones12} find that the dust mass-loss rate has a greater influence on the crystalline fraction than the gas mass-loss rate. This would indicate that the annealing is more important than the gas phase condensation, but the uncertainties in the gas mass-loss rates precluded a firm conclusion. Heating of the dust above the glass temperature could also be caused by shocks in the circumstellar environment. It is unclear how such a process could play a role in the outflow of single stars, but shocks caused by binary interaction could create the conditions for heating up the amorphous grains sufficiently \citep{edgar08} either in the outflowing material or in a circumbinary disk. In the case of disks, it would also be possible that the grains reside in the inner part of the disk close to the central star where the temperature is high enough for annealing \citep{gail01,ruyterThesis}. Crystalline silicates are, however, also known to exist in disks that are too cold for annealing \citep{molster99}. This suggests that it should also be possible to anneal amorphous grains in-situ in colder parts of the disk by shocks or other processes \citep{desch00, abraham09, edgar08}.

In this article we present Herschel/PACS spectra of 48 evolved stars in the 68-72~$\mu$m range. The sample contains stars in different post-main-sequence phases of evolution, among which are AGB (OH/IR) stars, different types of post-AGB stars, planetary nebulae, and a few massive ($>$10 $\text{M}_{\odot}$) evolved stars. The goal of this article is to (1) present the Herschel/PACS observations taken in three guaranteed and open-time programmes, together with the data reduction and (2) make a first analysis of Herschel/PACS data and determine the composition of the crystalline olivine in these sources. Subsequent publications will study the different subgroups of the sample in more detail. To study the forsterite content, including the bands at mid-infrared wavelengths, needs detailed radiative treansfer modelling because of the extreme high-density conditions in most of our sources. This would, however, be beyond the scope of this paper. A first paper by \cite{devries14} describes the results for the OH/IR stars. In Sect.~\ref{sec: data} we start with the sample selection and data reduction. In Sect.~\ref{sec: fitting} we describe how we fit and analyse the 69~$\mu$m bands in our sample. Our sample is described in more detail in Sect.~\ref{sec: resultsH4}, along with the results from fitting the 69~$\mu$m bands. We look at the laboratory studies of the 69~$\mu$m band in more detail in Sect.~\ref{sec: lab} and end this article with a discussion and conclusions in Sects.~\ref{sec: discussionH4} and \ref{sec: conclusionsH4}. The Appendices contain plots for all our reduced spectra.

\section{Sample selection and data reduction} 
\label{sec: data}

To study the formation and evolution of forsterite dust, we selected stars from different late evolutionary stages, as well as stars of intermediate and high initial masses (below and above approximately 8 $\text{M}_{\odot}$, respectively). We aimed at obtaining representative groups of stars for different classes: high mass-loss rate AGB stars (OH/IR stars), post AGB stars with and without circumbinary disks, planetary nebulae, massive evolved stars as luminous blue variables (LBV), and hyper giants, all known to contain forsterite dust.  
An initial selection of targets was obtained from a literature search. For almost all 26 initial targets, the crystalline oxygen-rich dust has already been observed through detection of mid-IR bands (11, 19, 23, 33, 40, or 43~$\mu$m), mostly in ISO-SWS spectra \citep{mol02_1}, but also from ground-based TIMMI spectroscopy  by \cite{ruyterThesis}. 

Herschel/PACS spectrometer observations \citep{pilbratt10, poglitsch10} of the targets were obtained in two guaranteed time observing programmes. The first was part of the "Mass loss of Evolved Stars (MESS)" \citep{groenewegen11_mess} guaranteed time key programme, for which Herschel/PACS "spectral energy distribution (SED)" 
mode spectra were taken covering the 55 to 200 $\mu$m range. The  selected sources which were not part of the spectrometer programme of MESS, were observed in the GT programme "Forsterite dust in the circumstellar environment of evolved stars" (GT1\_jblommae\_1). Because the latter programme focussed on detecting the 69~$\mu$m band, spectra were taken in the narrower wavelength range of 67-72 $\mu$m using the so-called "range scan" spectroscopy mode. For both the SED and range scan observations, the grating stepsize corresponded to Nyquist sampling (for further details, see \citealt{poglitsch10}). The spectral resolution around 70~$\mu$m is almost 1800 ($\lambda / \Delta \lambda$).  A typical RMS sensitivity of 1~Jy per resolution bin is obtained in the case of a single range scan measurement, which we used for all our sources with F$_{60\mu{\rm m}}$ above 100~Jy. We improved the S/N by increasing the number of repetitions for sources with lower fluxes. The intermediate flux stars (50 - 100~Jy) were observed with two repetitions (RMS sensitivity 0.7~Jy). Between 20 and 50~Jy, a repetition factor of 3 was used (RMS sensitivity 0.5~Jy) and for the faintest sources around 20~Jy we use four repetitions. 

To extend our sample and get more representatives for the different classes, we had a follow-up open time programme "Study of the cool forsterite dust around evolved stars" (OT2\_jblommae\_2). Here the selection was performed on the basis of the ISO-SWS catalogue by \cite{Kraemer02}.  All evolved stars that showed evidence of crystalline silicate emission bands in their SWS spectra (class SC or SEC), as well as sources with the 10~$\mu$m amorphous silicate band in (self-) absorption (SB and SA) and which were not part of the two earlier programmes, were added to the sample. 

We also included the spectra of the planetary nebula NGC~6543 and of OH~127.8+0.0, both obtained as a Herschel/PACS calibration measurement. The SED observation of NGC~6543 was reduced in the same way as the other sources in our programme. The spectrum of OH~127.8+0.0 was a combination of three SED measurements (obsids, 1342189956 up to 1342189958), and its data reduction is discussed in \citet{lombaert13}.

This adds up to a total of 48 spectra. Most of our sources are known to contain crystalline olivine dust through the detection of mid-infrared bands. Exceptions are NGC~6720 , AFGL~2019, AFGL~2298, and V432 Car. NGC~6720 was included as an additional oxygen-rich PN. AFGL~2019 is an OH/IR  star with the 10~$\mu$m amorphous silicate band in self-absorption, but no clear detection of forsterite bands in its SWS spectrum. Because of the low number of massive stars, we included AFGL~2298 and V432 Car. All sources are listed in Table \ref{tab: sources}. The sample contains 15 intermediate mass evolved stars with high mass-loss rates ($>5\cdot10^{-5}\,\text{M}_{\odot}\,\text{yr}^{-1}$), one AGB star with mixed chemistry (IRAS~09425$-$6040),  16 post-AGB stars with and without circumbinary disks, one D' type symbiotic star (HDE~330036), eight planetary nebulae, and seven massive evolved stars like (post-)RSGs (red super giants).

The spectra were reduced in the Herschel Interactive Processing Environment (HIPE, \citep{ott10}) package version 9, using the Herschel/PACS pipe-line script (ChopNodRangeScan.py). The absolute flux calibration used the PACS internal calibration block measurements, and for the spectral shape the PACS relative spectral response function was applied. The PACS integral-field-spectrometer contains 5$\times$5 spatial pixels, the so-called spaxels. The point spread function (PSF) of Herschel is larger than the central spaxel ($9.4^{\prime\prime} \times 9.4^{\prime\prime}$), and a correction needs to be made for the missing part of the PSF, even when combining several spaxels, to obtain the absolute source flux. Because of the pointing accuracy and jitter, it is best to combine as many spaxels under the PSF as possible. However, to keep the noise as low as possible, we combined all spaxels with a S/N higher than 10 rather than blindly adding the central 3$\times$3 spaxels. The S/N of every spaxel is obtained by calculating the standard deviation from a linear fit to the wavelength region of 67.5 to 68.0 $\mu$m and dividing this standard deviation by the signal. The sources AFGL 5379 and OH~21.5+0.5 were pointed off-target. For OH~21.5+0.5, all of the flux was still contained within the 5$\times$5 spaxels, and a spectrum could be extracted. The maximum flux within the spaxels for AFGL 5379 was found on one of the border spaxels, and thus not all flux of this object was detected so the flux levels should be taken as a lower limits. We flux-calibrated the spectra by calculating the total continuum flux (in the range of 67.5 to 68.0 $\mu$m) in the spectrum of all 3$\times$3 spaxels combined and then corrected this for the missing part of the PSF (correction factor 1.09, see PACS reduction manual). This flux value was used to scale the combined spaxels with S/N$>10$ to the proper continuum flux level.  

\section{Fitting the 69~$\mu$m bands}
\label{sec: fitting}

\begin{figure}
        \resizebox{\hsize}{!}{\includegraphics{\absPath  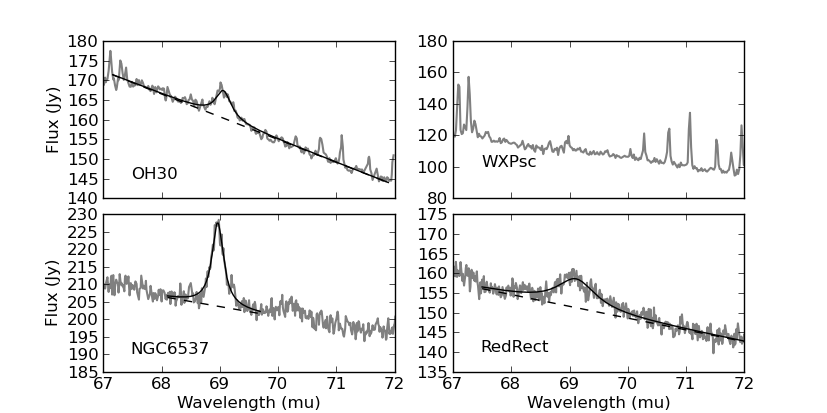}}
        \caption{ \textit{ Example spectra in grey with fitted Lorentzian in solid black and the continuum in dashed black. The shown sources are the OH/IR star OH~30.1$-0.7$ and WX Psc, the planetary nebula NGC 6537 and the circumbinary disk of the Red Rectangle. }}
        \label{fig: example69}
        \end{figure}

We measure the width and position of the 69~$\mu$m bands by making a Lorentzian fit to the band in the region of 67$-$72 $\mu$m. The observations are also simultaneously fitted with a slightly curved continuum (second-degree polynomial) as a continuum. The borders of the fitting region have been chosen differently in a few cases (see Appendix~A) because of the presence of structure in the continuum around the 69~$\mu m$ band. Most of the OH/IR stars and a few of the other objects in the sample show gas lines in their spectra (see Fig.~\ref{fig: example69}). The gas lines come from species like ortho- and para-H$_2$O, $^{12}$CO, $^{28}$SiO, and SO$_2$ \citep{decin12}. For sources with gas lines, we excluded the following wavelength regions in our fits: 66.9 -- 67.15; 67.22 -- 67.41; 67.47 -- 67.6; 68.91 -- 69.02; 70.6 -- 70.8; 71.0 -- 71.15; 71.5 -- 71.6; and 71.9 -- 72.2.

\begin{figure}
        \resizebox{\hsize}{!}{\includegraphics{\absPath  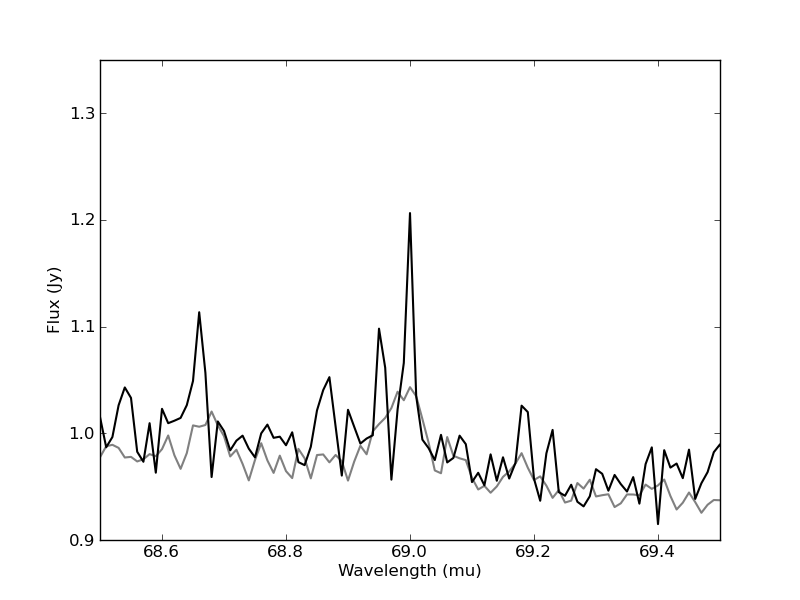}}
        \caption{ \textit{ Herschel/PACS spectrum of NML Tau with low (grey) and high (black) spectral resolution. The black spectrum with its higher spectral resolution shows that the gas line at 69~$\mu$m is a blend of two gas lines. This blend can not be recognized in the lower resolution spectrum in grey.}}
        \label{fig: NMLTau}
        \end{figure}

There is a blend of gas lines in the 67--72 $\mu$m wavelength range that is situated on top of the 69~$\mu$m band. At the spectral resolution of the PACS spectrometer in B2A ($\approx$ 1800 at 70~$\mu$m), this blend of gas lines shows as quite a broad feature and could be wrongly identified as a very weak and narrow 69~$\mu$m band of crystalline olivine. 
This feature can be resolved in spectra with a higher spectral resolution. For the star NML Tau, both a B2A and B3A (resolving power $\approx 5000$) spectrum was observed in the MESS programme \citep{groenewegen11_mess} (obsids: 1342203679 and 1342203681). In Fig.~\ref{fig: NMLTau}, a 1~$\mu$m range of the two spectra around 69~$\mu$m is shown. NML Tau contains no crystalline olivine, and the comparison of the two spectra indeed shows that the 69~$\mu$m feature is made up of at least two gas lines.  Since the lines occur in the spectra of our OH/IR stars, the wavelength region of 68.91 -- 69.02 is among the regions that were excluded from the forsterite band fit.

Errors on the width and position of the 69~$\mu$m band are obtained by a Monte Carlo method. The error on the properties of the 69~$\mu$m band are calculated by repeating the fitting of the band multiple times, while randomly varying the wavelength (by $\pm$0.02 $\mu$m from the standard value) of the borders of the fitting region and the wavelength of the gas lines that are excluded. The standard deviation of the sample of fitting results is then taken as the error. The width and position of the 69~$\mu$m band are plotted in Fig.~\ref{fig: pw_overview} and the width, position, and strength of the 69~$\mu$m band are listed in Table~\ref{tab: fitParameters}. All the fits and spectra are shown in the Figs.~\ref{fig: spec69_AGB} to \ref{fig: spec69_mass} in Appendix~A.
 A couple of weak broad features are seen in some of the spectra (as for NGC~6537 at 70.1~$\mu$m). None of these fitted the profile of crystalline olivine and have not been included further in our present analysis. In a future study we will search for solid state features in a wider spectral range than what is presented in this paper.

\cite{bowey02} performed a similar analysis of the ISO-LWS spectra of evolved stars. The sensitivity and spectral resolution of ISO-LWS was much lower (about 290 at 69~$\mu$m) than our data, limiting the determination of the width of the forsterite band. \cite{bowey02} used Gaussian fitting so that their results are not directly comparable to ours, especially with respect to the width of the feature. Despite these differences, we do find a similar range of peak wavelengths (68.92 -- 69.25~$\mu$m) for the forsterite band. 

\section{Results}
\label{sec: resultsH4}
        \begin{figure*}
        \resizebox{\hsize}{!}{\includegraphics{\absPath  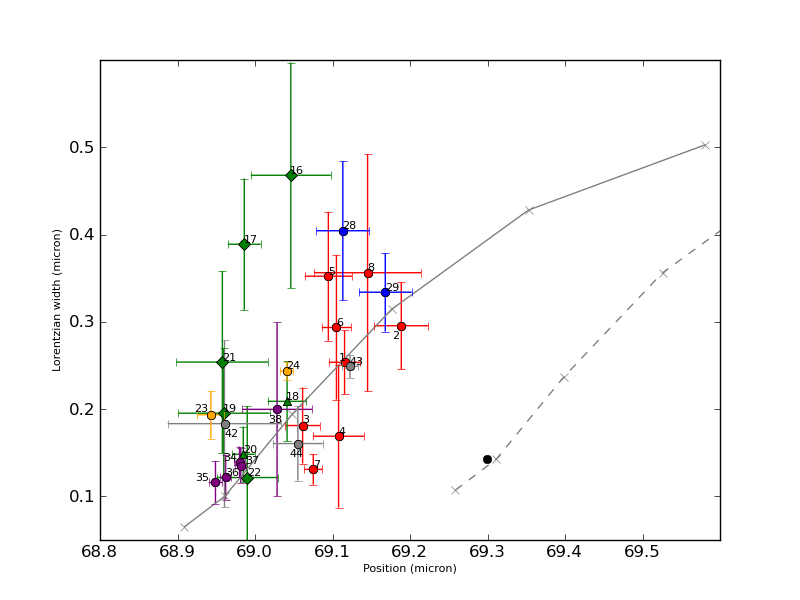}}
        \caption{ \textit{ {\bf Position and width of all detected 69~$\mu$m bands. In red are the OH/IR stars, in green the post-AGB stars, in blue the post-AGB stars with circumbinary disks, in yellow two other likely disk sources, in purple the PNs, and in grey the massive evolved stars. In contrast to the 69~$\mu$m bands of the evolved stars, the position and width of the 69~$\mu$m band of the disk around the young main-sequence star $\beta$ Pictoris is shown in black. In solid and dashed grey are the width and position of 69~$\mu$m bands using the laboratory measurements of \citet{koike03} and \citet{suto06}. The solid grey is for crystalline olivine without iron and the dashed grey with 1\% iron. The crosses on the curves indicate the temperatures going from 50, 100, 150, 200, and 295~K with increasing wavelength. The dashed grey curve is created using the extrapolation of \citet{devries12}}. 
}}
        \label{fig: pw_overview}
        \end{figure*} 

For almost all our targets, forsterite dust was already detected through the presence of mid-infrared bands. As mentioned in Sect.~1, the 69~$\mu$m band represents only one of in a total of 35 infrared active modes of crystalline olivine \citep{suto06} – the most prominent bands being located in the mid-infrared, namely at 11.3, 19.5, 23.6, and 33.6~$\mu$m. The precise band positions and profiles depend on the particle compositions, sizes, temperatures, and shapes.
A direct comparison of the strengths and positions of all forsterite bands detected in our sources is not possible because most
of our sample stars that show the 69~$\mu$m band in emission have optical depths of their dust shells larger than one in parts of the 10--50~$\mu$m wavelength range; e.g., for some of our OH/IR stars, the 10~$\mu$m optical depths reach values $\tau_{10}$ $>$ 15 \citep{suh13}. Therefore, deriving the forsterite temperature and mass fraction by an analysis of all observable mid- and far-infrared bands would require a large set of radiative transfer calculations, exceeding the scope of the present paper.
Furthermore, the mid-infrared features of olivine are less sensitive to the Fe content and temperature of the grains than the 69~$\mu$m band, which is an additional reason for focussing on the analysis of the latter in subsequent sections.

Figure~\ref{fig: example69} shows four example spectra in the range of the 69~$\mu$m band. The spectrum of OH~30.1$-0.7$ shows a strong 69~$\mu$m band, together with narrow gas lines. WX Psc has no 69~$\mu$m band, but shows the same strong gas lines as OH~30$.1-0.7$. Some of the strongest 69~$\mu$m bands are seen in planetary nebulae like NGC 6537. The Red Rectangle has one of the broadest 69~$\mu$m bands in the sample.

In Fig.~\ref{fig: pw_overview} the width and wavelength position of the detected 69~$\mu$m bands are plotted against each other. This figure also shows the width and position of laboratory measurements of crystalline olivine (explained in Sect.~\ref{sec: lab}). What can be seen from the figure is that all detected 69~$\mu$m bands have a width and position comparable to or above the curve of pure magnesium-rich crystalline olivine. In Sect.~\ref{sec: lab} we show that these widths and positions can only be explained by a pure magnesium-rich composition of the crystalline olivine and not by other effects. 

To discuss the sample further we subdivided it into four groups:
\begin{itemize}
\item 15 OH/IR stars with circumstellar outflows,
\item 18 post-AGB stars and other evolved objects,
\item 8 planetary nebulae,
\item  7 massive evolved stars.
\end{itemize}

\beginthetable
\centering
\begin{tabular}{ | l| l| l| l| l| l| l| l|}
  \hline
  Source name   &       IRAS                    &       \#      &pr     MM        &       Type                            &       Morphology              &       69 $\mu$m  & OBSID \\
  \hline \hline
 OH 32.8$-$0.3          &       18498$-$0017    &       1       &1              &       AGB                                 &       Outflow$^{1,2,3}$       &       +               &         1342209738 \\
 OH 21.5+0.5    &       18257$-$1000    &       2       &2              &       AGB                                 &       Outflow$^{1,3}$ &       +               & 1342242438      \\
 OH 30.1-0.7    &       18460$-$0254    &       3       &1              &       AGB                                 &       Outflow$^{2,3}$ &       +               & 1342216207      \\
 OH 26.5+0.6    &       18348$-$0526    &       4       &3              &       AGB                                 &       Outflow$^{1,2,3,4,5,6}$&        +       & 1342207777              \\
 OH 127.8+0.0   &       01304+6211      &       5       &               &       AGB                                 &       Outflow$^{1,2,3}$       &       +               & {\it combined, see text}        \\
 AFGL 2403      &       19283+1944      &       6       &2              &       AGB                                 &       Outflow$^{7,8}$ &       +               & 1342245226      \\
 IRAS 21554     &       21554+6204      &       7       &2              &       AGB                                 &       Outflow$^{7,9}$ &       +               & 1342247150      \\
 AFGL 5379      &       17411$-$3154    &       8       &3              &       AGB                                 &       Outflow$^{3, 4, 7}$     &       +       &       1342228537      \\
 OH 104.9+2.4   &       22177+5936      &       9       &1              &       AGB                                 &       Outflow$^{2,3,10,11}$&  -       &         1342212261      \\
 AFGL 4259      &       20043+2653      &       10      &2              &       AGB                                 &       Outflow                 &       -       &       1342244918      \\
 IRAS 17010     &       17010$-$3840    &       11      &1              &       AGB                                 &       Outflow$^{2,12}$        &       -       &       1342216176      \\
 RAFGL 2374     &       19192+0922      &       12      &2              &       AGB                                 &       Outflow$^{12,13}$       &       -       &       1342244920      \\
 WX Psc                 &       01037+1219      &       13      &3              &       AGB                                 &       Outflow$^{4,14, 15}$&   -       &       1342202122      \\
 IRC +50137     &       05073+5248      &       14      &2              &       AGB                                 &       Outflow$^{7, 13,15}$&   -       &       1342249315      \\
 AFGL 2019      &       17501$-$2656    &       15      &2              &       AGB                             &       .                               &       -       &       1342252253      \\
  \hline \hline
 RAFGL 4205     &       14562$-$5406    &       16      &2      &       C/O pAGB/pPN                &       Multipolar$^{20}$        D               &       +       &       1342249211      \\
 AFGL 6815      &       17150$-$3224    &       17      &3      &       pAGB/pPN                &       Bipolar/rings$^{21,22}$   D       &       +       &       1342216629      \\
 HD 161796      &       17436+5003      &       18      &3      &       pAGB/pPN                 &       Torus/bipolar$^{16,17}$   S       &       +               &       1342208882 \\
 IRAS 16342     &       16342$-$3814    &       19      &3      &       pPN                              &       Torus/Bipolar$^{18,19}$   D       &       +               & 1342216628      \\
 CPD $-$64 2939 &       14331$-$6435    &       20      &2      &       pAGB/pPN                &       .$^{23,24}$                       S               &       +               &       1342250006 \\
 OH 231.8+4.2   &       07399$-$1435    &       21      &3      &       pAGB/pPN                &       Bipolar/binary$^{25,26}$        D       &       + &       1342196695              \\
 IRAS 16279     &       16279$-$4757    &       22      &3      &       C/O pAGB                    &       Bipolar$^{26,27}$               D       &       +       &       1342228204      \\
   HDE 330036   &       15476$-$4836    &       23      &1      &       Symbiotic                       &       Shells/binary$^{28}$            &       +       & 1342225842              \\
 IRAS 09425     &       09425$-$6040    &       24      &3      &       C/O AGB                     &       disk?$^{36,40-42}$                 &       +       &         1342225564      \\
IRAS 19306      &       19306+1407      &       25      &2      &       pAGB/pPN                &       Bipolar$^{29,30}$               D       &       -               & 1342244921      \\
 HD 331319      &       19475+3119      &       26      &2      &       pAGB/pPN                &       Multipolar/torus        $^{31}$ D       &       -       &       1342245232      \\
 Hen 2$-$90             &       13064$-$6103    &       27      &2      &       pAGB/pPN                 &       Torus/binary$^{32}$             D       &       -       &       1342248308      \\
 \hline
 Red Rectangle &        06176$-$1036    &       28      &3      &       pAGB                         &       circumbinary disk$^{33-36}$     &       + &               1342220928      \\
 ACHer          &       18281+2149      &       29      &1      &       pAGB                         &       circumbinary disk$^{37-39}$     &       +       &       1342208896      \\
 AR Pup         &       08011$-$3627    &       30      &1      &       RV Tauri                   &       circumbinary disk$^{35}$                &       -       &       1342210381      \\
 HD 93662               &       10456$-$5712    &       31      &1      &       RV Tauri                   &       circumbinary disk$^{35}$                &       - &               1342211823      \\
 IW Car         &       09256$-$6324    &       32      &1      &       RV Tauri                   &       circumbinary disk$^{35}$                &       -       &         1342210386      \\
 U Mon          &       07284$-$0940    &       33      &1      &       RV Tauri                   &       circumbinary disk$^{35}$                &       - &       1342206993              \\
 \hline 
\hline
 NGC 6543       &       17584+6638      &       34      &4      &       PN                               &       Bipolar$^{43,44}$                               &       +       &                1342187203 \\
 NGC 6302       &       17103$-$3702    &       35      &3      &       PN                               &       Bipolar/torus$^{45,46}$         &       +       &       1342230151      \\
 NGC 6537       &       18021$-$1950    &       36      &3      &       PN                               &       Bipolar, shells$^{47}$                  &       +       &       1342231323      \\
 CPD --56$^\circ $8032 &        17047$-$5650    &       37      &3      &       C/O PN                      &       Bipolar, torus/disk $^{26,56}$  &       +       &       1342228202      \\
 BD +30 3639    &       19327+3024      &       38      &3      &       PN                               &       Ring$^{50-52}$                       &       +       &       1342220601      \\
 IRAS07027      &       07027$-$7934    &       39      &2      &       PN                               &       .$^{53,54}$                             &       -       &       1342245248      \\
 IRAS17347      &       17347$-$3139    &       40      &3      &       C/O PN                      &       Bipolar/torus$^{30,55}$         &       -       &       1342229697      \\
 NGC 6720       &       18517+3257      &       41      &1      &       PN                               &       Bipolar, rings$^{48,49}$                &       ?       &       1342208920      \\
  \hline \hline
AFGL 4106       &       10215$-$5916    &       42      &1              &       post-RSG                        &       Oval/knots$^{26,57}$                      &       +       & 1342207818              \\
HD 168625       &       18184$-$1623    &       43      &3              &       pAGB/LBV?               &       Bipolar/torus$^{58,59}$           &       +       & 1342229741              \\
AFGL 2343       &       19114+0002      &       44      &3              &       p-AGB/p-RSG             &       Spherical shell$^{26}$             &       +       & 1342208894            \\
AFGL 2298       &       18576+0341      &       45      &1              &       Massive                  &       Knotty shell$^{60}$                      &       -       & 1342207804              \\
IRC +10420      &       19244+1115      &       46      &3              &       p-RSG                    &       Bipolar$^{61}$                           &       -       & 1342208886              \\
NML Cyg         &       -                       &       47      &3              &       Variable                         &       Asymmetric$^{62}$                        &       -       & 1342198175              \\
V432 Car                &       11065$-$6026    &       48      &1              &       LBV                      &       Bipolar$^{63}$                           &       -       & 1342212229              \\
\hline
\end{tabular}
 \caption{ List of the observed stars (in order of OH/IR stars, subsequently post-AGB and others and the PN, and  finally the massive evolved stars), their IRAS numbers, observation programme, types, morphologies and whether the 69~$\mu$m band was detected in the spectrum. The observation programme numbers refer to: 1) "Forsterite dust in the circumstellar environment of evolved stars", 2) "Study of the cool forsterite dust around evolved stars", 3) MESS, 4) "Mapping the distribution of the crystalline silicate forsterite in the Cat's Eye Nebula (NGC 6543)". References are: (1) \cite{suh13}, (2) \cite{devries10}, (3) \cite{just06}, (4) \cite{just12}, (5) \cite{ches05}, (6) \cite{groenewegen12}, (7) \cite{beck10}, (8) \cite{begeman97}, (9) \cite{kemper03}, (10) \cite{riechers04}, (11) \cite{riechers05}, (12) \cite{kwok97}, (13) \cite{just92}, (14) \cite{decin07}, (15) \cite{suh02}, (16) \cite{hoogzaad02}, (17) \cite{min13}, (18) \cite{verhoelst09}, (19) \cite{dijkstra03}, (20) \cite{sahai11}, (21) \cite{ramoslarios12}, (22) \cite{kwok98}, (23) \cite{suarez06}, (24) \cite{gauba04}, (25) \cite{sanchezcontreras04}, (26) \cite{lagadec11}, (27) \cite{matsuura04}, (28) \cite{angeloni07}, (29) \cite{cerrigone09}, (30) \cite{sahai07}, (31) \cite{MingChienHsu11}, (32) \cite{Sahai00}, (33) \cite{waters98nat}, (34) \cite{bujarrabal05}, (35) \cite{ruyter06}, (36) \cite{molster01}, (37) \cite{vanwinckel98}, (38) \cite{giridhar00}, (39) \cite{jura99}, (40) \cite{molster99nat},  (41) \cite{mol02_3}, (42) \cite{hernandez06}, (43) \cite{waters96}, (44) \cite{balick04}, (45) \cite{matsuura05_6302}, (46) \cite{kemper02_6302}, (47) \cite{matsuura05_6537}, (48) \cite{odell07}, (49) \cite{hoof10}, (50) \cite{gillett73}, (51) \cite{phillips07}, (52) \cite{waters98}, (53) \cite{demarco02}, (54) \cite{zijlstra91}, (55) \cite{gregoriomonsalvo04},  (56) \cite{cohen99}, (57) \cite{molster99},  (58) \cite{umana10}, (59) \cite{smith07}, (60) \cite{buemi10},
 (61) \cite{oudmaijer13}, (62) \cite{monnier97}, (63) \cite{riera95} }
\label{tab: sources}

\endthetable

\subsection{OH/IR stars}
The first group is composed of oxygen-rich AGB stars (see Table~\ref{tab: sources}) that are currently in a state of high mass-loss (called the superwind, $>5\cdot10^{-5}\,\text{M}_{\odot}\,\text{yr}^{-1}$, \citealt{bedijn87}). These sources are called OH/IR stars because of their maser emission, and they are only visible in the infrared because the central star is completely enshrouded by its optically thick dusty outflows. The mass-loss rates of the OH/IR sources in our sample range from relatively low to high. For example, the object WX Psc is at the lower end with a mass-loss rate of $0.6\cdot10^{-5}\,\text{M}_{\odot}\,\text{yr}^{-1}$ \citep{decin07}, while sources such as OH~26.5+0.6, OH~127.8+0.0, OH~30.1$-0.7,$ and OH~32.8$-0.3$ have mass-loss rates all the way up to $(15-20)\cdot10^{-5}\,\text{M}_{\odot}\,\text{yr}^{-1}$ \citep{just96, just92, schut89, groene94}. Most of the  OH/IR stars in our sample are situated close to the galactic plane and have high luminosities (a few 10,000~L$_{\odot}$) and long-period variability ($>1000$ days) \citep{devries14} 
so are expected to  have initial masses in the range of $3-8\,\text{M}_{\odot}$ \citep{baud81}.
The high density and the oxygen-rich composition of the outflow of these OH/IR stars leads to efficient condensation of crystalline olivine \citep{waters96, tielens98}. The abundance of crystalline olivine in these outflows can be as high as 12 \% \citep{devries10}. 

For eight out of the fifteen OH/IR stars, we detected the 69~$\mu$m band (see Fig.~\ref{fig: spec69_AGB} in Appendix~A). The strongest bands are for OH~32$.8-0.3$ and IRAS 21554, which have a peak strength that is $\sim$8\% of the continuum. The weakest bands are seen for OH~26.5+0.6 and AFGL 5379, which have a peak strength that is $\sim$3 \% of the continuum. Almost all sources in the group of OH/IR stars show signs of gas lines in their spectrum except for IRAS 21554 and IRAS 17010.

All detected 69~$\mu$m bands have a width and position close to those of crystalline olivine with zero percent iron and grain temperatures of 100-200 K, with an average of $\sim$170 K (see Fig.~\ref{fig: pw_overview}). Some of the OH/IR stars have 69~$\mu$m bands that are wider than the pure magnesium-rich crystalline olivine bands. This is an indication of a temperature gradient in the outflow, as is explained further in Sect.~\ref{sec: lab}.

\subsection{Post-AGB stars and other low-mass evolved objects}
The second group in our sample is a collection of 16 post-AGB sources and two other evolved low-mass objects. Out of the eighteen stars, we detect the 69~$\mu$m band in eleven objects. Unlike AGB stars, which show predominately spherical symmetric circumstellar shells, post-AGB stars, as do planetary nebulae, contain aspherical structures, such as outflowing tori, disks, and/or bipolar outflows. Table~\ref{tab: sources} gives an overview of the different classes of morphologies that are observed in our sample. 

A first sub-group that we discuss is the sample of six binary post-AGB stars with confirmed Keplerian disks (see Table~\ref{tab: sources}). The dust processing is known to be strong in these disks, resulting in large grain sizes and high crystallinity \citep{molster99nat,gielen08,gielen11}. The Red Rectangle among them is a well known object that has a circumbinary disk that is viewed edge-on and which completely obscures the central binary (\cite{cohen04}). The gas in the disk of the Red Rectangle is known to rotate with Keplerian velocities (\cite{bujarrabal05}) and the gas disk contains about $(2-6)\,\times10^{-3}\,\text{M}_{\odot}$ (\cite{bujarrabal05}).

For the Red Rectangle and also AC Her, the dust in the disk is known to be partly crystalline, and indeed for both we detect a 69~$\mu$m band. The peak over continuum strength of the 69~$\mu$m bands of the Red Rectangle and ACHer are about 5-6\% of the continuum, which is comparable to the other post-AGB stars without confirmed Keplerian disks and the OH/IR stars. The remaining four disk sources, AR Pup, HD 93662, U Mon, and IW Car, were taken from a study by  \cite{ruyter05}. These stars are known to contain crystalline olivine from their spectral features in the 8-13~$\mu$m range. From those features \cite{ruyter05} was not able to derive abundances of crystalline olivine compared to the continuum dust species, but it must be in the range $\sim$10-50\%. Strikingly, we do not detect a 69~$\mu$m band for these objects. This can be explained by the presence of hot ($\sim 600$~K)  dust as will be discussed further in Sect.~\ref{sec: discussionH4}.

The widths and positions of the 69~$\mu$m band of the binary post-AGB stars with Keplerian disks are shown in blue, while the other post-AGB stars are shown in green in Fig.~\ref{fig: pw_overview}. Most of the post-AGB stars have 69~$\mu$m bands that are located slightly above the curve of pure forsterite in Fig.~\ref{fig: pw_overview}. As will be shown in Sect.~\ref{sec: lab}, this is probably caused by the presence of a temperature gradient in the circumstellar environment. 

Two other sources that are likely to contain disks are IRAS~09425 and HDE~330036. The first, an example of a mixed chemistry object, is a \text{J-type} carbon AGB star.  Its circumstellar environment shows signs of both carbon and oxygen-rich gas and dust species, and it is extremely crystalline. \cite{molster01} find a crystallinity of 75\% by mass for IRAS~09425. It was the record holder for a long time until \cite{jiang13} reported a crystallinity fraction above 90\% for the PN IRAS~16456--3542. The high crystallinity of IRAS~09425 can be best understood if one assumes that the silicates are part of a (circumbinary) disk. \cite{molster99nat} show that this assumption is further strengthened by the high flux ratio of mm-wave over 60~$\mu$m observed in IRAS~09425, which indicates the presence of large grains, which are most likely formed by coagulation in a long-lived circumstellar disk. There is, however, no direct evidence of, the disk structure to date and the debate on the nature of IRAS~09425 is still open. Garcia-Hernandez et al. (2006) conclude that the mixed chemistry is best explained by a recent change of oxygen- to carbon-rich chemistry after a dredge-up. It is, however, difficult to understand such a scenario considering the extremely high crystallinity. IRAS~09425 also has the strongest 69~$\mu$m band in our sample, with a peak strength of 29\% of the continuum. HDE~330036 is a D' type symbiotic star, the evolutionary stage following the post-AGB (Jorissen et al 2005). Crystalline olivine bands were detected at 19.7, 23.7, and 33.6~$\mu$m,  and pyroxenes at 15.9, 20.7, and 26.1~$\mu$m in the ISO-SWS spectrum by \cite{angeloni07}. They find that the crystalline silicates reside in a circumbinary disk at a temperature close to 100 -- 200~K. Our 69~$\mu$m detection of the forsterite is relatively strong with a peak strentgh of 8.5\% over the continuum. From the width versus peak position diagram (Fig.~\ref{fig: pw_overview}), we find a temperature in agreement with the 100--200~K range by  \cite{angeloni07}. As for the other disk sources, it is positioned above the pure forsterite curve. 

Of the remaining sample of ten post-AGB stars, seven have a detection of the 69~$\mu$m band. The average peak over continuum strength is about 4.3 \%, similar to the OH/IR stars but lower than for the planetary nebulae. On basis of an optical imaging survey of reflection nebulae around post-AGB stars, \cite{ueta00} concluded that all post-AGB stars with detected nebulosity are aspherical. The types of asphericity could be divided in two classes that depended on the central star obscuration.  The SOLE class ("star obvious low-level elongated") is interpreted to contain an optically thin torus around the central star. The DUPLEX type ("dust prominent longitudinally extended") contains blobs on opposite sides and has an optically thick torus that hides the central star in optical wavelengths. This division of classes was studied further and confirmed in studies of the mid-infrared morphologies \citep{ueta02, meixner02}. \cite{ueta00} show that the two classes can be separated on the basis of infrared colours. For the ten remaining post-AGB stars in our sample, we indicate to what class they belong on the basis of 2MASS J$-$K$_s$ colour. SOLE and DUPLEX sources clearly occupy different parts in the width versus peak position diagram (Fig.~\ref{fig: pw_overview}). All DUPLEX sources fall above the curve of laboratory measurements of pure forsterite. As  discussed in Sect.~\ref{sec: discussionH4}, the extreme widths could be related to a temperature gradient possibly in a  torus-like structure around these objects (Sect.~\ref{sec: lab}). 

The SOLE sources CPD~-642939 and HD161796 fall close to the pure forsterite curve and fall in between the classes of OH/IR stars and PNe. HD161796 is a well-studied post-AGB star that is believed to have evolved directly from an OH/IR star progenitor. This source has an oxygen-rich envelope containing about 4\% crystalline olivine (\cite{hoogzaad02})  and also has a
peak strength of its 69~$\mu$m band of 4\% of the continuum. The mass loss of HD161796 stopped roughly 300 years ago (\cite{min13}), and during the hundreds of years of active mass loss that followed, the mass-loss rate of HD161796 increased from $\sim7\cdot10^{-5}\,\text{M}_{\odot}\,\text{yr}^{-1}$ up to almost $\sim50\cdot10^{-5}\,\text{M}_{\odot}\,\text{yr}^{-1}$ (\cite{min13}).

\subsection{Planetary nebulae}
The third group in our sample consists of planetary nebulae. All sources except NGC 6720 
and IRAS 17347 show strong crystalline olivine bands in the ISO-SWS spectra. For NGC 6720, both the ISO-SWS and our Herschel-PACS spectrum may have missed the presence of crystalline silicates because the dusty ring structure of NGC 6720 ($>45^{\prime\prime}$, \cite{hoof10}) falls mostly outside the field-of-view of both apertures ($50^{\prime\prime} \times 50^{\prime\prime}$). However, a visual inspection of the LWS spectrum (aperture $\approx 80^{\prime\prime}$) also did not reveal any forsterite dust at 69~$\mu$m.  The ISO-SWS spectra of IRAS17347 shows no spectral features of crystalline olivine, and we also detect no 69~$\mu$m band. \cite{cohen02} claim detecting forsterite bands at 23.7, 33.7, and 69~$\mu$m for IRAS07027. Our PACS spectrum does not confirm their 2.5~$\sigma$ detection of the 69~$\mu$m band. For the detected 69~$\mu$m bands, we find that the peak over continuum strengths  is high and close to 10\% of the continuum. Only BD~+303639 has a weaker 69~$\mu$m band of 4\% of the continuum.

All planetary nebulae in this sample for which fosterite is detected contain very high density regions, 
such as a torus a disk or a shell consisting of high density knots. In some cases, such as NGC 6302 and NGC 6537, the very dense and massive central torus or disk almost completely obscures the central star. \cite{matsuura05} find that the central obscuration in NGC~6537 can be explained by a high mass loss at the end of the AGB.
Our detected sources all have bipolar outflows with dense stellar winds with high velocities of at least 200 km/s.
They have rather high central star core masses and are thought to come from higher mass stars (3--8 M$_\odot$).  

BD+303639 [WC9], CPD --56$^\circ $8032 [WC10], and IRAS07027 are member of the class of planetary nebulae with very low-excitation Wolf-Rayet spectra. \cite{cohen02} studied the infrared spectra of these sources and found evidence of dual chemistry through the presence of PAHs and amorphous and crystalline silicates.  For the objects IRAS07027 and BD~+303639, it is suggested that the AGB star experienced a final thermal pulse before leaving the AGB phase, which caused the inner parts of the dust envelope to be carbon rich (\cite{waters98, matsumoto08, zijlstra91}). It is not clear if the progenitor AGB star of all such nebula evolved from an oxygen-rich to a carbon-rich AGB star. As is the case for post-AGB stars (e.g. the Red Rectangle), the dual chemistry can also be explained by the presence of the silicates residing in a (circumbinary) disk. IRAS~07027 \citep{tLH91} and IRAS~17347 \citep{zijlstra89} are also are detected as OH maser sources, indicating that these sources only recently became planetary nebulae and evolved from the class of OH/IR stars. Remarkably, only IRAS~07027 shows evidence of crystalline silicates and neither of the two has a 69~$\mu$m detection.  

The five planetary nebulae with detected 69~$\mu$m bands all have widths and positions that are very close to those of pure forsterite.  On average the crystalline olivine in the PNe has a temperature of around 100~K.

\beginthetable
\centering
\begin{tabular}{|l |c |r |r |r |r |}
  \hline
  Source name   &       \#      & A (Jy) & A/cont & $\Delta$p ($\times$10) & w ($\times$10) \\
                        &               &  & ($\times$100) & $\mu m$ & $\mu m$ \\
  \hline \hline
  
OH~32.8$-0.3$ & 1 & 5.2$\pm$0.4 & 7.2$\pm$0.6 & 1.2$\pm$0.2 & 2.5$\pm$0.4 \\ 
OH~21.5+0.5 & 2 & 4.4$\pm$0.3 & 5.5$\pm$0.4 & 1.9$\pm$0.3 & 3.0$\pm$0.5 \\ 
OH~30.1$-$0.7 & 3 & 7.2$\pm$1.0 & 4.5$\pm$0.6 & 0.6$\pm$0.2 & 1.8$\pm$0.4 \\ 
OH~26.5+0.6 & 4 & 13.3$\pm$2.7 & 3.0$\pm$0.6 & 1.1$\pm$0.3 & 1.7$\pm$0.8 \\ 
OH~127.8+0.0 & 5 & 4.5$\pm$0.3 & 4.6$\pm$0.3 & 0.9$\pm$0.3 & 3.5$\pm$0.7 \\ 
AFGL2403 & 6 & 4.8$\pm$0.4 & 5.9$\pm$0.5 & 1.1$\pm$0.2 & 2.9$\pm$0.8 \\ 
IRAS21554 & 7 & 2.8$\pm$0.3 & 8.2$\pm$0.9 & 0.8$\pm$0.1 & 1.3$\pm$0.2 \\ 
AFGL5379 & 8 & 16.0$\pm$3.8 & 2.7$\pm$0.6 & 1.5$\pm$0.7 & 3.6$\pm$1.4 \\ 
\hline 
RAFGL4205 & 16 & 5.5$\pm$0.9 & 3.4$\pm$0.6 & 0.5$\pm$0.5 & 4.7$\pm$1.3 \\ 
AFGL6815 & 17 & 10.5$\pm$0.8 & 4.9$\pm$0.4 & -0.1$\pm$0.2 & 3.9$\pm$0.8 \\ 
HD161796 & 18 & 5.3$\pm$1.1 & 4.3$\pm$0.9 & 0.4$\pm$0.2 & 2.1$\pm$0.5 \\ 
IRAS16342 & 19 & 6.2$\pm$3.2 & 1.9$\pm$1.0 & -0.4$\pm$0.6 & 2.0$\pm$0.7 \\ 
CPD-642939 & 20 & 5.2$\pm$0.8 & 8.8$\pm$1.3 & -0.2$\pm$0.1 & 1.5$\pm$0.3 \\ 
OH~231.8+4.2 & 21 & 26.9$\pm$4.6 & 2.6$\pm$0.4 & -0.4$\pm$0.6 & 2.5$\pm$1.0 \\ 
IRAS16279 & 22 & 5.5$\pm$1.5 & 4.1$\pm$1.1 & -0.1$\pm$0.4 & 1.2$\pm$0.8 \\ 
HDE330036 & 23 & 1.4$\pm$0.1 & 8.5$\pm$0.7 & -0.6$\pm$0.2 & 1.9$\pm$0.3 \\ 
IRAS09425 & 24 & 6.2$\pm$0.2 & 28.6$\pm$0.9 & 0.4$\pm$0.1 & 2.4$\pm$0.1 \\ 
RedRect & 28 & 7.3$\pm$0.6 & 4.8$\pm$0.4 & 1.1$\pm$0.3 & 4.0$\pm$0.8 \\ 
ACHer & 29 & 1.4$\pm$0.1 & 6.0$\pm$0.4 & 1.7$\pm$0.3 & 3.3$\pm$0.5 \\ 
\hline 
NGC6543 & 34 & 9.5$\pm$1.6 & 10.2$\pm$1.7 & -0.2$\pm$0.1 & 1.4$\pm$0.2 \\ 
NGC6302 & 35 & 93.4$\pm$16.9 & 10.3$\pm$1.9 & -0.5$\pm$0.1 & 1.2$\pm$0.2 \\ 
NGC6537 & 36 & 24.1$\pm$3.9 & 11.9$\pm$1.9 & -0.4$\pm$0.1 & 1.2$\pm$0.3 \\ 
CPD-568032 & 37 & 18.0$\pm$4.4 & 9.3$\pm$2.3 & -0.2$\pm$0.1 & 1.4$\pm$0.2 \\ 
BD~+303639 & 38 & 5.4$\pm$1.6 & 4.0$\pm$1.2 & 0.3$\pm$0.5 & 2.0$\pm$1.0 \\ 
\hline 
AFGL4106 & 42 & 11.2$\pm$4.3 & 1.7$\pm$0.7 & -0.4$\pm$0.7 & 1.8$\pm$1.0 \\ 
HD168625 & 43 & 11.1$\pm$0.4 & 13.1$\pm$0.5 & 1.2$\pm$0.1 & 2.5$\pm$0.1 \\ 
AFGL2343 & 44 & 16.7$\pm$3.2 & 3.6$\pm$0.7 & 0.6$\pm$0.3 & 1.6$\pm$0.4 \\ 
\hline 

\end{tabular}
 \caption{ Best fit parameters for the Lorentz fits to the detected 69 $\mu m$ bands: amplitude (A), wavelength shift of the peak wavelength position relative to 69.0 $\mu m$ ($\Delta$p), and the width of the Lorentz curve (w).   }
\label{tab: fitParameters}
\endthetable

\subsection{Massive stars}
The last and fourth group contains eight evolved stars that have a massive ($>$8 $\text{M}_{\odot}$) progenitor.
Three of these sources have a detected 69~$\mu$m band. All the circumstellar environments of these objects show mild to strong deviations from spherical symmetry. For the source HD168625, it is not clear that this object is a post-AGB star or a luminous blue variable. AFGL 4106 is a confirmed binary (\cite{molster99}). For AFGL 2298 and V432 Car we do not find any 69~$\mu$m band, and indeed the ISO-SWS spectra also do not show any indication of crystalline olivine. On the other hand, we also do not find a 69~$\mu$m band for  IRC +10420 and NML Cyg even though these sources show a weak 33.6 $\mu$m band of crystalline olivine in their ISO-SWS spectra.

The width and position of the three detections are shown in grey in Fig.~\ref{fig: pw_overview} and the temperatures range from $\sim$50-150 K. The strength of the band varies from 1.7 to 3.6 and 13.1 \% of the continuum for AFGL 4106, HD 168625, and AFGL 2343, respectively.


\section{Laboratory measurements of the 69 $\mu$m band.}
\label{sec: lab}
Using the available laboratory measurements, we now describe the different parameters that have an effect on the width and position of the 69 $\mu$m band. We describe the effects of the grain temperature, iron content of the mineral, grain shape, grain size, and the presence of a temperature gradient.

        \begin{figure*}
        \resizebox{\hsize}{!}{\includegraphics{\absPath  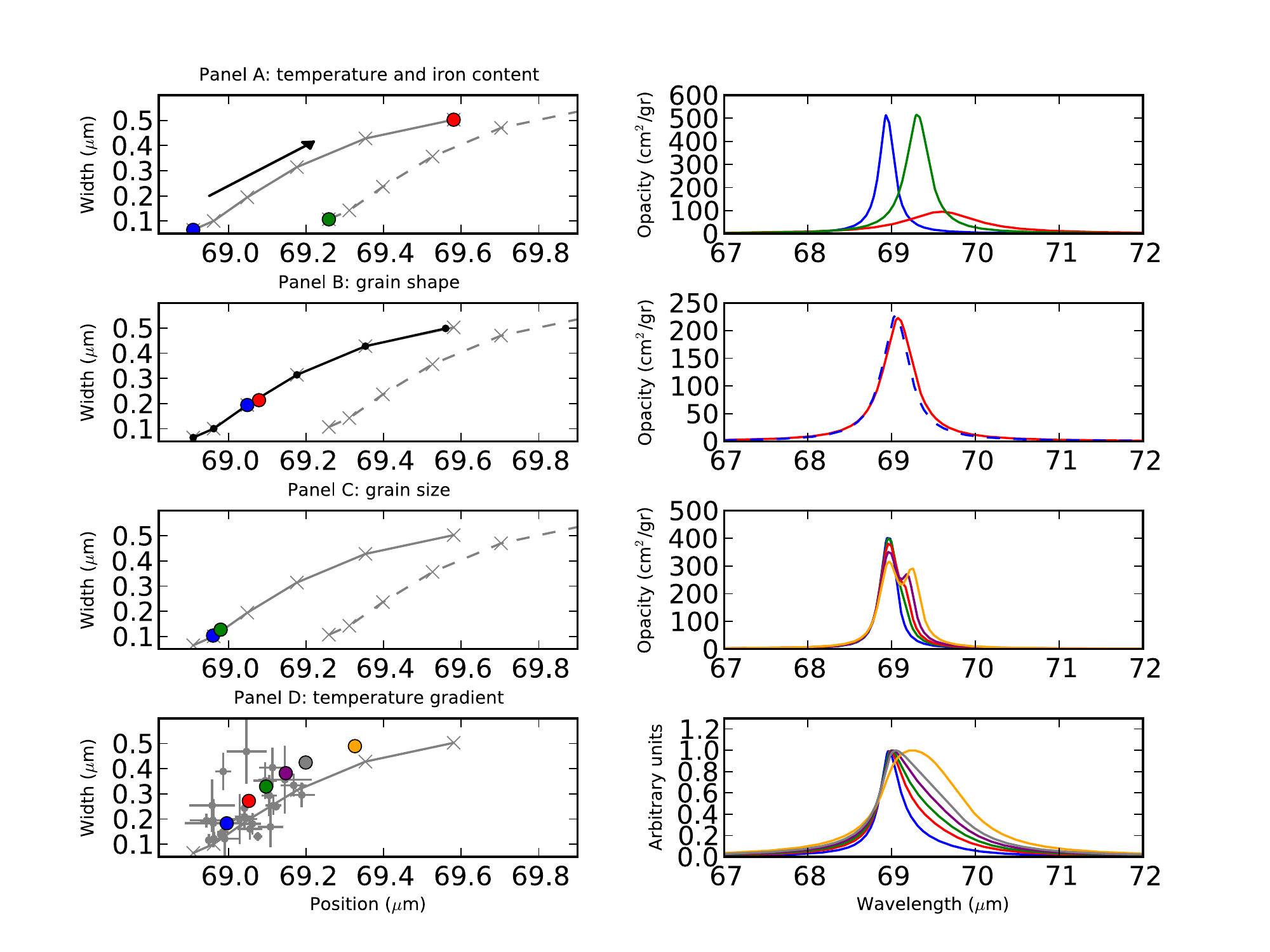}}
        \caption{The influences of different parameters on the width and position of the  69 $\mu$m band as described in Sect.~\ref{sec: lab}.
In the left column the (Lorentzian) width is plotted against the wavelength position of the 69 $\mu$m band. In the right column examples of the absorption opacities for the 69 $\mu$m band are shown. The colours of the bands on the right correspond to the coloured dots in the figures on the left. The solid and dashed grey curves in panel A are as in Fig. \ref{fig: pw_overview}.The black arrow in panel A indicates the direction of the increasing temperature of the laboratory measurements (from 50 to 295 K). Panel B shows in solid black the width and position of the 69 $\mu$m band of crystalline olivine with 0\% iron and calculated with DHS. The blue dashed band corresponds to DHS and the red one to CDE. Panel C shows the position and width of the 69 $\mu$m band for different grain sizes (at T = 100 K). The 69 $\mu$m bands are plotted for the grain sizes 1 (blue), 4 (green), 5 (red), 6 (purple), and 7 $\mu$m (orange). Panel D shows the width and position of a 69 $\mu$m band coming from a spherically symmetric distribution of crystalline olivine. The dots and bands with the colours blue, red, green, purple, orange, and grey have an ($i_{\rho}$, $i_{\text{T}}$) of (2.0,~0.5), (2.0,~1.0), (2.0,~2.0), (3.0,~0.5), (4.0,~0.5), and (4.0,~2.0), respectively (see Sect. \ref{sec: lab}). The observed 69 $\mu$m bands are also shown in grey with error bars. The 69~$\mu$m bands coming from a temperature gradient shown in panel D vary strongly in peak strength. To clearly show the broadening of these bands we normalized the bands to one.}
        \label{fig: pw_lab}
        \end{figure*} 

\subsection{The 69 $\mu$m band of crystalline olivine at a single temperature}

\cite{suto06} have measured the refractive indices of crystalline olivine at the temperatures of 50, 100, 150, 200, and 295~K. Panel A of Fig. \ref{fig: pw_lab} shows the width and position of the measurements of \cite{suto06}. They made reflection measurements, and we converted the optical constants they obtained into opacities. To make this conversion, we assumed the grain shape distribution CDE (continuous distribution of ellipsoids, \cite{BH83}). From Fig. \ref{fig: pw_lab}A it can be seen that the 50 K measurement has the bluest wavelength position and the most narrow 69 $\mu$m band. The band broadens and shifts to the red as the grain temperature increases. 

\subsection{The 69 $\mu$m band of crystalline olivine containing iron}
\cite{koike03} have measured the absorbance of crystalline olivine for different iron contents of the crystals. The step size in iron content in their measurements is large: after the 0\% iron measurements, the first measurement is at $\sim8$\%. \cite{koike03} only did these measurements at room temperature. We use the interpolations made by \cite{devries12} to generate 69 $\mu$m bands of crystalline olivine at different compositions and temperatures. This interpolation assumes that the shift due to the iron content of the crystal is the same at all temperatures. Panel A shows how the width and position of the 69 $\mu$m band shifts when the iron content of the crystals is increased by 1\%. Due to an iron increase of 1\%, the band's wavelength position shifts to the red by more than $\sim$0.2~$\mu$m. 

\subsection{The 69 $\mu$m band and grain properties}
\label{sec: sizeshape} 
In this section we test the effect of grain shape and size on the width and position of the 69 $\mu$m band. We calculate opacities from the optical constants of \cite{suto06} using the grain shape distribution CDE (\cite{BH83}) and DHS (distribution of hollow spheres, \cite{min03}, filling factor 0.8). In the case of DHS we also test different particle sizes (0.01-7.0~$\mu$m). We cannot test the particle size for CDE, because this size distribution is limited to particle sizes that are much smaller than the wavelength. We do not consider spherical grains, since \cite{min03} have shown that the use of homogeneous spheres as particle shapes introduces unrealistic effects, due to resonances that are introduced by the symmetry of these particles.

Panel B of Fig. \ref{fig: pw_lab} shows the difference in width and position of the 69 $\mu$m band between the grain shape distributions CDE and DHS (at grain size 0.01 $\mu$m). The width and position of the 69 $\mu$m band for grains with a DHS shape distribution are slightly more narrow and bluer, respectively, but the difference is insignificant.

For the DHS shape distribution, panel C of Fig. \ref{fig: pw_lab} shows how the 69 $\mu$m band of a 100 K grain changes as a function of grain size. The shape of the 69~$\mu$m band for grain sizes smaller than 1~$\mu$m (not shown) are the same as for 1~$\mu$m. For grain sizes between 1-4~$\mu$m there is a small but insignificant effect on the width and position of the 69 $\mu$m band. Going to grain sizes larger than 4 $\mu$m, the band develops a double-peaked shape, which is not recognized in the observations.

\subsection{Crystalline olivine with a temperature gradient}
The width and position of the 69 $\mu$m band can be made more narrow or red-shifted compared to the width and position of crystalline olivine with 0\% iron by including iron. Here we show that it is also possible to have 69 $\mu$m bands that are broader and/or bluer than those of crystalline olivine without including iron. This is possible by introducing a temperature gradient. A temperature gradient will create a 69 $\mu$m band that is a combination of 69 $\mu$m bands of different temperatures. We do not want to consider the full parameter space accessible by introducing a temperature gradient, but simply illustrate the consequences of such a gradient with an example. We assume a spherically symmetric density distribution where the density goes as $\rho = \rho_{0}\,r^{-i_{\rho}}$, where $r$ is the distance to the centre of the distribution. Furthermore, we assume the temperature of the crystalline olivine to follow $T=T_{\text{in}}\,r^{-i_{\text{T}}}$, where $T_{\text{in}}$ is the maximum temperature at the inner starting point of the distribution. Rearranging factors, we can write
\begin{equation}
r = \left( \frac{T}{T_{\text{in}}} \right)^{-1/i_{\text{T}}}
\end{equation}
\begin{equation}
\text{d}r = -\frac{1}{i_{\text{T}} \, T_{\text{in}}} \left( \frac{T}{T_{\text{in}}} \right)^{-\frac{1}{i_{\text{T}}}-1} \text{d}T
.\end{equation}
The flux coming from crystalline olivine with such a temperature gradient will, in the optically thin limit, be proportional to
\begin{equation}
F \propto \int_{50\,\text{K}}^{300\,\text{K}} \frac{1}{ i_{\text{T}} \, T_{\text{in}} } \left( \frac{T}{T_{\text{in}}} \right) ^{ \frac{i_{\rho}-i_{\text{T}}-3} {i_{\text{T}}}} \text{d}T
.\end{equation}
The numerical integration is done over a grid of temperatures with a step size of 25 K. Band shapes for temperatures in between temperatures at which laboratory measurements are available are generated using the interpolations of \cite{devries12}. A typical density and temperature gradient for a constant mass-loss outflow around an optically thick OH/IR star is $i_{\rho}=2$ and $i_{T}=0.5$ (\cite{devries10}). In Panel D of Fig. \ref{fig: pw_lab}, we show the result of this calculation for ($i_{\rho}$, $i_{\text{T}}$) equal to (2.0, 0.5), (2.0, 1.0), (2.0, 2.0), (3.0,~0.5), (4.0,~0.5), and (4.0,~2.0). The results in panel D of Fig. \ref{fig: pw_lab} show that by introducing a temperature gradient, the width of the 69 $\mu$m band can be broader than those of crystalline olivine olivine with 0\% iron at a single temperature.

\section{Discussion} 
\label{sec: discussionH4}

\subsection{Width and peak position of the 69~$\mu$m band}
From the width and position diagram in Fig.~\ref{fig: pw_overview}, it is clear that all evolved stars have 69~$\mu$m bands with a position and width that can only be explained by pure forsterite. Only in a few cases, such as OH~26.5+0.6 and IRAS21554, maybe 0.5\% iron can be present in the crystalline olivine. In the past, the ISO and Spitzer spectra of evolved stars have shown no spectral features of pure fayalite. 
Investigation of the infrared bands in the ISO and Spitzer spectra showed  an upper limit of the iron content of crystalline olivine of 10~\%  (\cite{tielens98, mol02_3}). Our study at a higher spectral resolution and sensitivity at 69~$\mu$m shows that the crystalline olivine only contains magnesium and no iron. 

For disks around young stellar objects, the 69~$\mu$m band has also been used to probe the composition and location of the olivine in the disk, showing that in these disks the crystalline olivine is also magnesium-rich, but may contain a few percent of iron ($<$0-3\% iron, \cite{mulders11, devries12, sturm13}). Figure~\ref{fig: pw_overview} also shows the 69~$\mu$m band of the disk around the young main-sequence star $\beta$ Pictoris \citep{devries12}. Even though the crystalline olivine in the disk of $\beta$ Pictoris only contains 1\% iron, the width and position of its 69~$\mu$m band shown in Fig.~\ref{fig: pw_overview} clearly contrast with those of the evolved stars. This contrast hints at a subtle difference between the crystalline olivine in disks around young stars and that in the circumstellar environments of evolved stars.

Figure~\ref{fig: pw_overview} shows that some sources have broader features than those of pure magnesium-rich crystalline olivine. This indicates that the emission of the crystalline olivine is not dominated by a single temperature, but comes from a circumstellar environment with a temperature gradient. For example, among the \text{OH/IR} stars OH~127.8+0.0 and AFGL 5379 and among the post-AGB stars the Red Rectangle, RAFGL 4205, OH~231.8+4.2, and AFGL 6815 have 69~$\mu$m bands broader than those of pure magnesium-rich crystalline olivine. In Fig.~\ref{fig: pw_lab} we show that most of these broad 69~$\mu$m bands can be explained by a temperature gradient. 

The post-AGB stars with such broad 69~$\mu$m bands are all classified as DUPLEX sources. These typically have very high-density tori or disks, causing high optical depths \citep{ueta00, meixner02}. The high optical depth possibly causes the temperature gradient in the circumstellar environment of these stars. Interestingly, the two SOLE objects (HD~161796 and CPD~-64 2939), which have much lower density environments and are optically thin, do not show such broad bands and fall close to the 0\% forsterite line in Fig.~\ref{fig: pw_overview}.

\subsection{The formation of forsterite}
 In this section we discuss scenarios that can explain the formation of pure magnesium-rich forsterite in the circumstellar environment.
A first scenario for forming crystalline olivine in the outflow of evolved stars is by direct condensation from the gas. Here we describe three possible reasons that iron is not incorporated in crystalline olivine. The first is that there is too little gaseous iron in order to condense iron bearing crystalline olivine. Because at thermal equilibrium metallic iron is stable at higher temperatures than crystalline olivine, it is expected that metallic iron condenses before crystalline olivine, severely lowering the partial pressure of iron. But when exactly metallic iron and crystalline olivine condense is very dependent on environmental properties. For example, \cite{woitke06_radpres} shows that metallic iron could condense at lower gas temperatures than crystalline olivine, because of the high opacity of metallic iron in the wavelength range ($\sim$1~$\mu$m) at which the central star of AGB stars mainly emits its energy. This means that for outflows of AGB stars, metallic iron is stable farther out from the central star than expected from thermal equilibrium calculations.

When gaseous iron is available for the formation of crystalline olivine, it will only be incorporated in the lattice if the partial pressure of iron becomes high compared to that of magnesium. A second reason could then be that the magnesium vapour pressure only drops low enough in the temperature and pressure regions where olivine does not form crystalline anymore \citep{gailsedl99} and thus that the crystalline silicates will only contain magnesium, 

which corresponds well with the composition that we find. Then, in the lower temperature regions where olivine forms in amorphous form, the vapour pressure of magnesium drops low enough and iron can be used for either condensation of amorphous olivine grains or metallic iron grains \citep{gailsedl99}. 
Even though we find that the crystalline olivine dust has very low iron content, this does not need to be the case for the amorphous form of olivine. \cite{suh99} finds optical constants for the amorphous silicates that resemble the experimental results from \cite{Dorsch95} for a Mg$_{0.8}$Fe$_{1.2}$SiO$_4$ composition. Modelling by \cite{devries10} of the spectrum of Mira gave a composition of Mg$_{1.36}$Fe$_{0.64}$SiO$_4$ for the amorphous silicate. Although the composition may not be accurately known, the amorphous silicates in the circumstellar environment of mass losing AGB stars are likely to contain a significant fraction of iron in the grain.

A third reason that iron would not be used for crystalline olivine is a kinetic. It is possible that the reactions in an outflowing environment will quickly freeze out due to the rapidly decreasing pressure. This could mean that before iron can be used in crystalline olivine, the density of the gas has dropped so low that the reactions responsible for the formation of fayalite do not take place.

Gas phase condensation is not the only possible formation mechanism of crystalline olivine. It is also possible to form crystalline olivine by annealing amorphous olivine. When amorphous olivine is heated above the glass temperature of olivine, it can rearrange its lattice elements into a crystalline form.  \cite{sogawa99} describe such a scenario for the outflow around evolved stars. In the case of a sufficiently high mass-loss rate ($> 3 \times 10^{-5} $ M$_\odot$ / yr for a $2 \times 10^4 $ L$_\odot$ AGB star), heterogeneous grains consisting of a corundum core and a silicate mantle can reach high enough temperatures for annealing.
In the case of circumstellar disks, it is possible that the grains reside in the inner part of the disk close to the central star where the temperature is warm enough for annealing. It is also possible to anneal amorphous grains in-situ in colder parts of the disk by shocks or other processes \citep{desch00, abraham09, edgar08}. The composition of crystalline olivine formed by these processes is difficult to predict, since it is unlikely that the in-situ formation of crystals happens under equilibrium conditions. The only laboratory measurements where amorphous olivine grains are annealed are by \cite{davoisne06}. They show that when an amorphous olivine material containing both magnesium and iron is annealed, it produces metallic iron inclusions and a magnesium-rich crystalline olivine material. This experiment shows that the magnesium-rich composition we find for crystalline olivine formed around evolved stars could also be explained by the annealing of amorphous grains.

\subsection{Comparison of the different classes}
\cite{devries14} studied the sub-class of OH/IR stars by comparing the forsterite bands, including the 69~$\mu$m one presented in this paper, with model spectra. They  argue that the most likely interpretation of the observed 69~$\mu$m bands of OH/IR stars is that the forsterite is formed in the outflow and that the forsterite temperature reflects the duration of the so-called superwind, a phase of intense mass loss ($\approx 10^{-4}$ M$_\odot$/yr). The OH/IR stars for which no 69~$\mu$m band is detected (seven out of fifteen in our sample) are explained by a superwind that only started recently (about a hundred years) and for which the forsterite dust is too warm and shows only mid-infrared bands. 
Cooler dust temperatures reflect longer duration superwinds, where the dust had the chance to travel farther away from the star. The coolest detected forsterite (at about 150~K) corresponds to a superwind radius of about 2500 AU or to a superwind duration of about 1200 years. The implications on the total mass loss and the AGB evolution are discussed further in \cite{devries14}. When the OH/IR star stops its mass-loss and evolves away from the AGB, the material that was ejected during the AGB will drift away and continue to cool down. An example of an object that has such a cooling dust shell is HD~161796. This source stopped its mass loss roughly 300 years ago (\cite{min13}). Figure \ref{fig: pw_overview} indeed shows that HD~161796 is positioned next to the OH/IR stars at a slightly lower temperature found for the coolest forsterite of OH/IR stars (150~K). Our sample also shows that the average temperature of the crystalline olivine gets colder when going from the AGB to the post-AGB stars and PNe ($\sim$170 K to $\sim$100 K). We note that our sample is not selected in a way that we can link these sources on an evolutionary basis, but this temperature decrease is consistent with the scenario of a gradually cooling circumstellar environment. Also, one should be careful about interpreting the temperatures indicated by the 69~$\mu$m band, since the dusty environments of these evolved stars are very likely optically thick and the emission from them depends on many properties of the system. Another limitation in this comparison is that for some of our sources, such as the Red Rectangle and AC Her, the forsterite is not part of an outflow but rather exists in a stable (circumbinary) disk.

The peak strength of the 69~$\mu$m band of the PNs is $\sim$10\% of the continuum, while those of the post-AGB and the OH/IR stars are between $\sim$2-8\% (see Table~\ref{tab: fitParameters}). The strength of these bands should be interpreted carefully because it can depend on properties of the system in the case of a high optical depth. But the stronger 69~$\mu$m bands of the PNs are very likely due to the crystalline olivine in these objects being colder, and for colder crystalline olivine the peak strength of the 69~$\mu$m band increases.

In comparison to the rest of our sample, IRAS09425 is exceptional with a band strength of 27\% of the continuum. This source is known to be extremely crystalline, and this high crystallinity can be understood best by the presence of a stable disk (\cite{molster01}). We must conclude that for IRAS09425 the formation mechanism of crystalline olivine and/or the circumstances under which it takes place are exceptional and probably different from the processes in any of our other objects. Then it is striking that the composition of the crystalline olivine in IRAS09425 is the same as for the other sources in our sample. 

Furthermore, if we compare the other objects with confirmed disks to IRAS09425, then the Red Rectangle and ACHer have a peak strength of the 69~$\mu$m band that is much smaller than IRAS09425 (5\%\ and 6\% compared to 27\%). And the other disk sources, AR Pup, HD~93662, and IW Car, do not even show a 69~$\mu$m band, while they clearly show crystalline olivine bands in the wavelength range of 8-13~$\mu$m (\cite{ruyterThesis}). This makes for the interesting situation where the crystalline olivine in these three disks is probably hot ($\sim$600 K) crystalline olivine dust. The opacity of forsterite at these temperatures no longer shows a 69~$\mu$m feature, while it still shows the spectral features at lower wavelengths. Indeed, \cite{ruyterThesis} finds that the temperature of the dust in the disks of AR Pup, HD~93662, and IW Car is $\sim$600 K and concludes that the flux comes from the inner parts of the disk. Our sample now marks three distinct kinds of disk sources. One is IRAS09425, which is extremely crystalline, but the crystalline olivine is relatively cold ($\sim$120 K). The Red Rectangle and AC Her have moderately strong 69~$\mu$m bands and crystalline olivine temperatures in the order of 200 K. In contrast AR Pup, HD~93662, and IW~Car show no 69~$\mu$m bands, but have hot crystalline olivine, which means the crystalline olivine is situated close to the central star.

\section{Conclusions} 
\label{sec: conclusionsH4}
We can summarize our conclusions as follows.
\begin{itemize}
\item The width and position of the 69~$\mu$m band shows that the crystalline olivine formed in outflows and disks around evolved stars is pure forsterite and contains no iron.
\item The temperature indicated by the 69~$\mu$m band is ${\sim100-200~\text{K}}$. On average the crystalline olivine in the OH/IR stars has the highest temperature, with decreasing temperatures over post-AGB to PNe.
\item  Eight of fifteen OH/IR stars show 69~$\mu$m band detection. \cite{devries14} show that the forsterite band properties can be explained by the formation of forsterite dust in the superwind. The non-detections in our sample correspond to sources where the superwind only started recently (about a hundred years), and the coolest temperatures correspond to the longest duration of the superwind, about 1200 years.
\item Most of the  post-AGB stars have 69~$\mu$m bands broader than that of pure magnesium-rich olivine, which is a sign of a temperature gradient in the disk or outflow. 
\item The sources with confirmed disks show distinct differences. The Red Rectangle and AC Her have moderate amounts of crystalline olivine at moderate temperatures. IRAS09425's 69~$\mu$m band confirms its extreme crystallinity and shows it is relatively cold. The objects AR Pup, HD93662, and IW Car have no detectable 69~$\mu$m band, which indicates that the crystalline olivine in their disks is hot ($\sim$600 K) and situated close to the central star.
\end{itemize}

\begin{acknowledgements} 
   JB, PR, BvB acknowledge support from the Belgian Science Policy Office through the ESA Prodex program.
   BdV acknowledges support from the Fund for Scientific Research of
   Flanders (FWO) for his Aspirant fellowship, as well as under grant number G.0470.07. 
  F.K. acknowledges funding by the Austrian Science Fund FWF under project number P23586. We also thank the anonymous 
  referee for the valuable comments.
\end{acknowledgements}

\bibliographystyle{aa}
\bibliography{references}

\def\appendixNameA{}
\def\appendixNameB{}

\appendix
\section{\appendixNameA 69 $\mu m$ bands and fits}

        \begin{figure*}
        \resizebox{\hsize}{!}{\includegraphics{\absPath  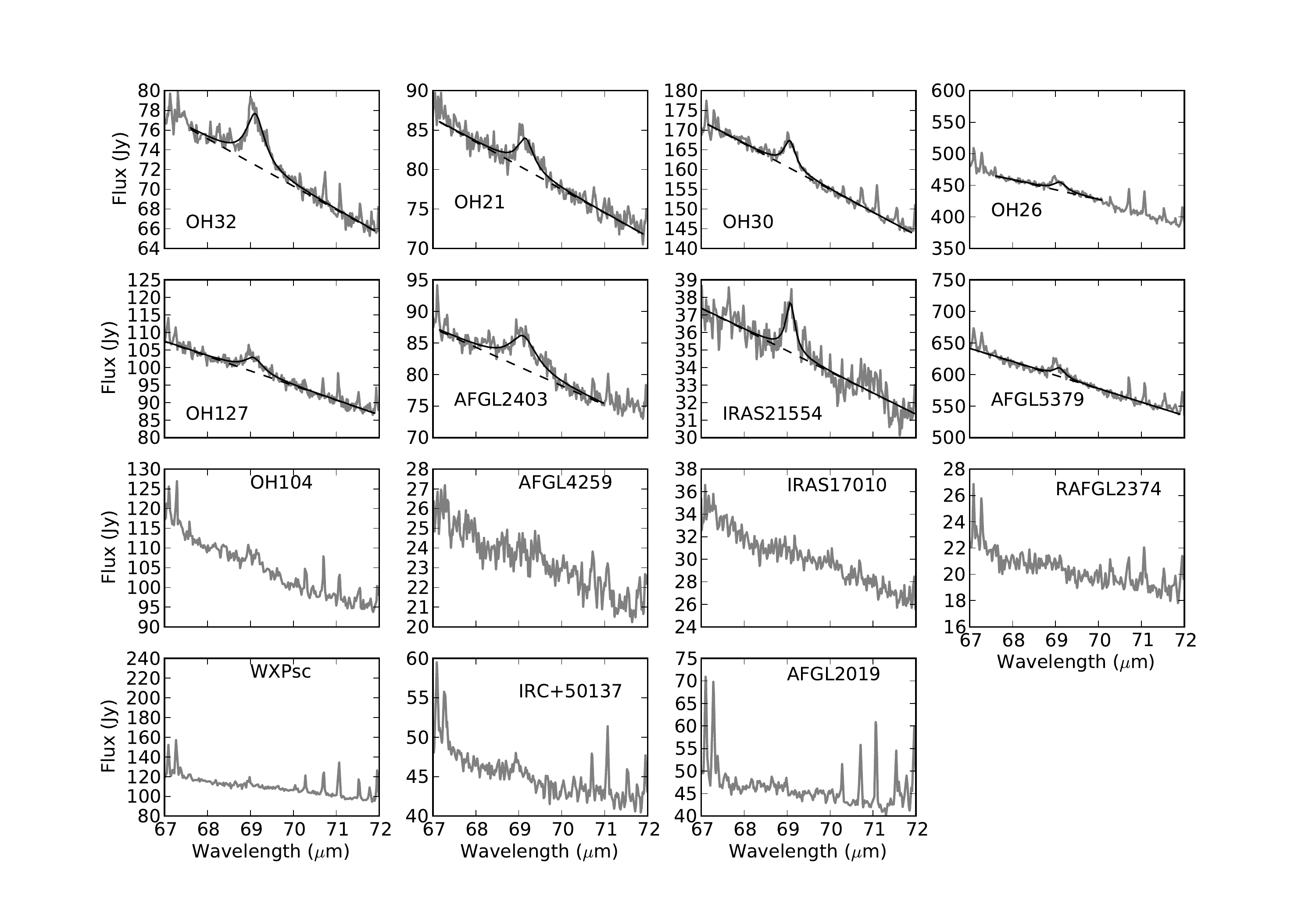}}
        \caption{ \textit{ Spectra of the OH/IR stars.  }}
        \label{fig: spec69_AGB}
        \end{figure*}

        \begin{figure*}
        \resizebox{\hsize}{!}{\includegraphics{\absPath  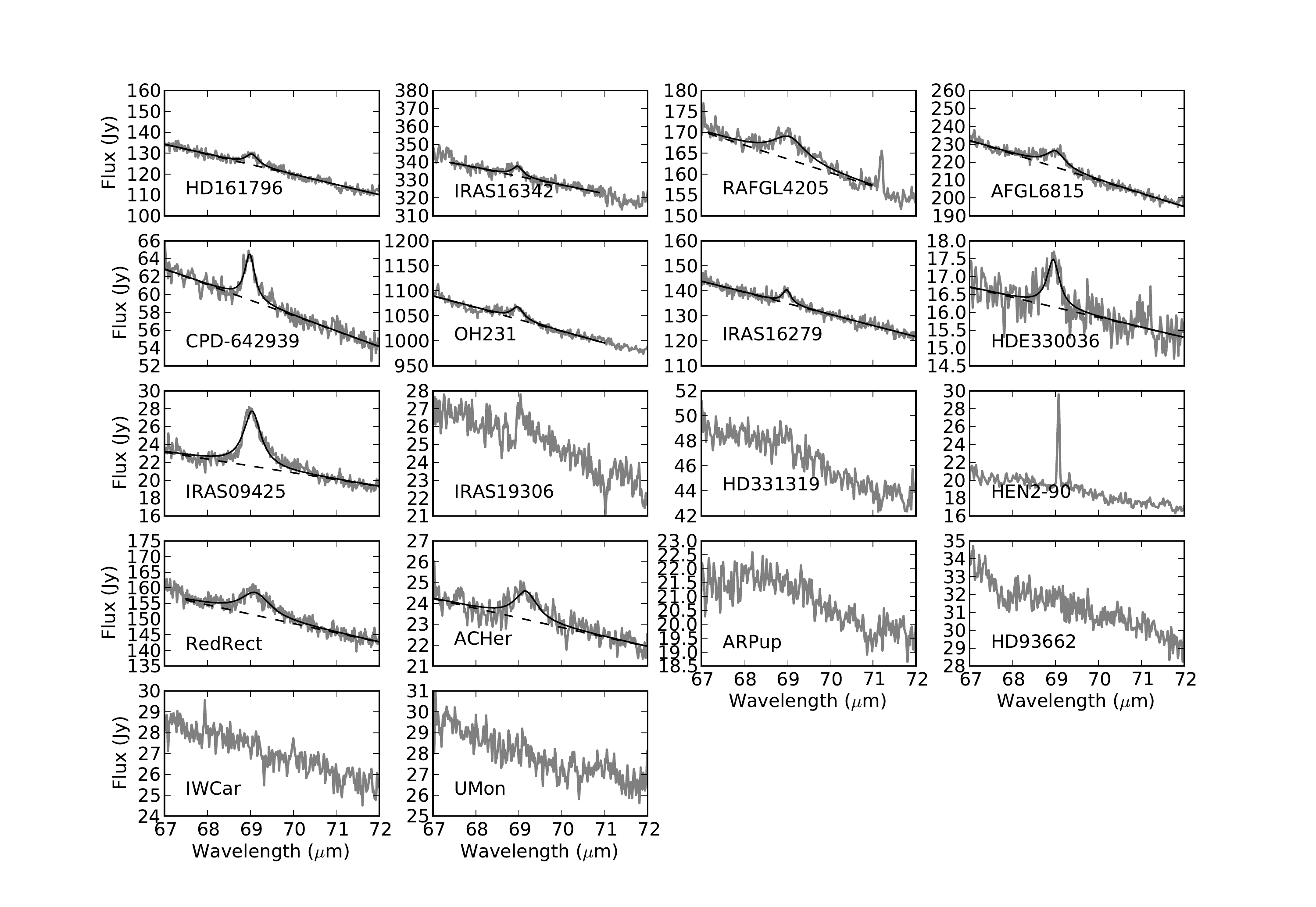}}
        \caption{ \textit{Spectra of post-AGB stars and other sources}}
        \label{fig: spec69_pAGB}
        \end{figure*}

        \begin{figure*}
        \resizebox{\hsize}{!}{\includegraphics{\absPath  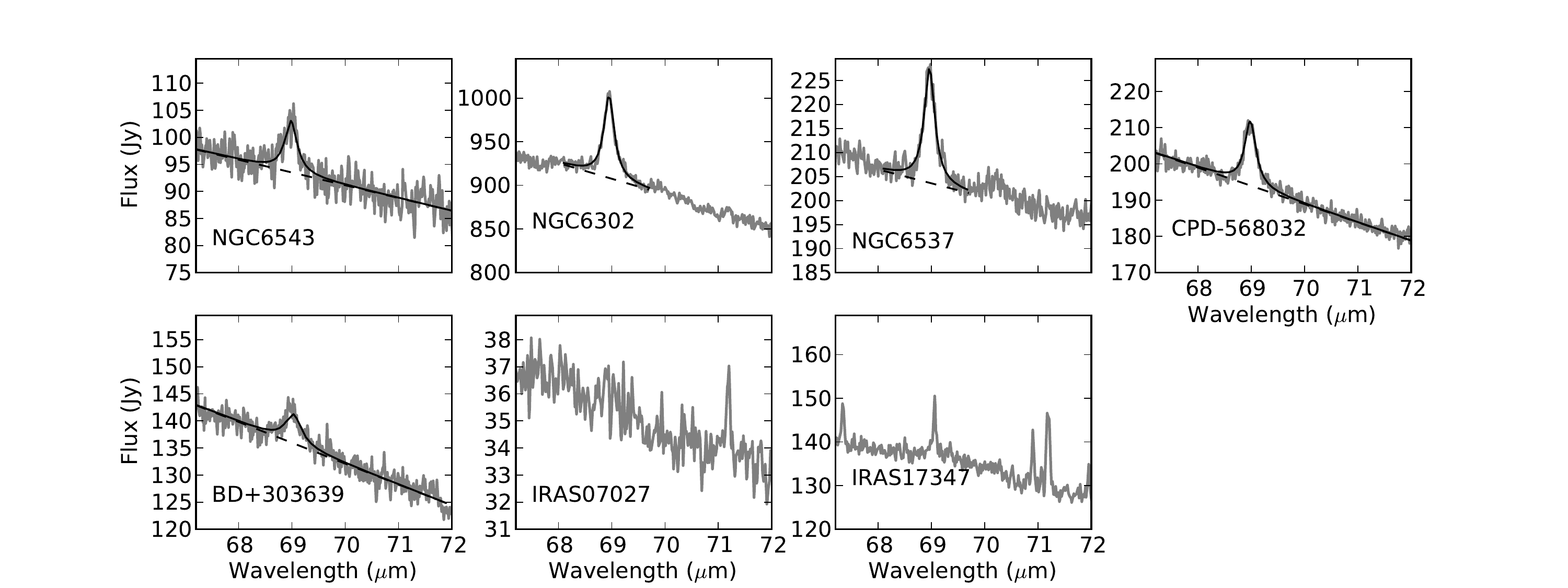}}
        \caption{ \textit{Spectra of PNe }}
        \label{fig: spec69_PN}
        \end{figure*}

        \begin{figure*}
        \resizebox{\hsize}{!}{\includegraphics{\absPath  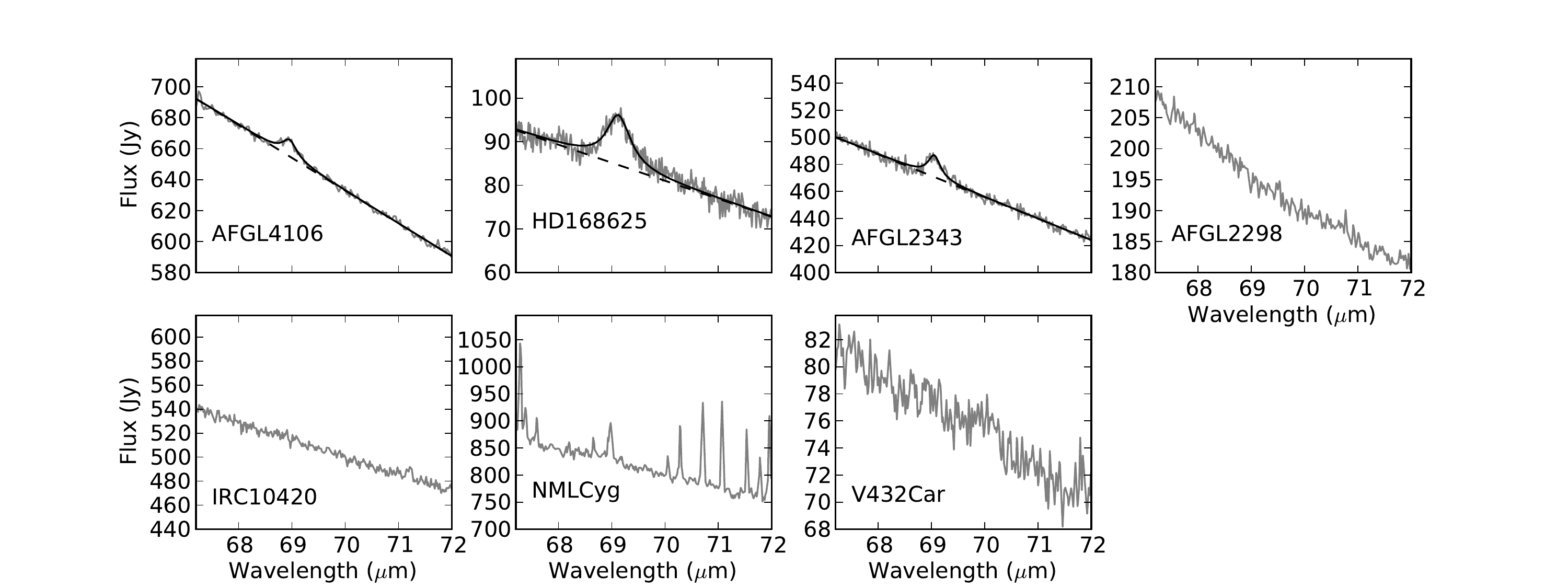}}
        \caption{ \textit{Spectra of the massive evolved stars }}
        \label{fig: spec69_mass}
        \end{figure*}

\end{document}